\documentclass[aps,prb,preprint, showpacs]{revtex4-1}

\usepackage{amsmath}
\usepackage{amssymb}
\usepackage{bm}
\usepackage{epsfig}
\usepackage{graphicx}
\usepackage{color}

\def\be{\begin{equation}}
\def\ee{\end{equation}}
\def\bea{\begin{eqnarray}}
\def\eea{\end{eqnarray}}

\begin{document}

\newcount\timehh  \newcount\timemm
\timehh=\time \divide\timehh by 60
\timemm=\time
\count255=\timehh\multiply\count255 by -60 \advance\timemm by \count255

\title{ Fine Structure of the band edge Excitons and  Trions  in CdSe/CdS Core/Shell Nanocrystals.}

\author{A. Shabaev}
\affiliation{George Mason University, Fairfax VA 22030, USA}

\author{A. V. Rodina}
\affiliation{Ioffe Physical-Technical Institute RAS, 194021 St.-Petersburg, Russia}

\author{Al. L.~Efros}
\affiliation{Naval Research Laboratory, Washington DC 20375, USA }

\begin{abstract}
We present a  theoretical description of excitons and positively and negatively charged trions in ''giant`` CdSe/CdS core-shell nanocrystals (NCs). 
 The developed theory provides the parameters describing  the fine structure of excitons in CdSe/CdS core/thick shell NCs  as a function of the CdSe/CdS conduction band offset and the CdSe core radius.   We have also developed a general theory describing the  fine structure of positively charged trions created in semiconductor NCs with a degenerate valence band. 
The calculations take into account the complex structure of the CdSe valence band and inter-particle Coulomb  and exchange interaction.  Presented in this paper are the  CdSe core size  and CdSe/CdS conduction band offset dependences (i) of  the positively charged trion fine structure, (ii) of  the binding energy of the negatively charged trion, and (iii) of the radiative decay time for excitons and trions. The  results of theoretical calculations are in qualitative agreement with available experimental data.
\end{abstract}

\maketitle

\section{introduction}

Growing attention to  CdSe/CdS core/thick shell nanocrystals (NCs) initially reported by  two groups \cite{MahlerNM08,ChenJACS08} is stimulated by  their superior  optical properties superior to any other  NC hetero-structures prepared up to now. The photoluminescence is never completely quenched  in these NCs \cite{SpinicelliPRL09} and  the blinking  is almost completely suppressed \cite{MahlerNM08,BrovelliNature11}.    At low temperatures these structures demonstrate suppression of non-radiative Auger recombination \cite{HtoonNL10} and almost 100\%  photoluminescence quantum yield \cite{ParkPRL11}. As a result these NC structures demonstrate a very low threshold for optically pumped lasing \cite{Garcia-SantamariaNL09}.

The origin of such outstanding optical properties of the  CdSe/CdS core/thick shell nano-structures (or giant CdSe/CdS NCs \cite{ChenJACS08}) is not completely understood, but one thing is clear. The very thick shell  always prevents carrier escape to the  NC, no matter which mechanism is responsible for this escape:  direct quantum mechanical tunneling, thermo-ionization of the NCs, or Auger  auto-ionization of the NCs, where two electron hole pairs were created simultaneously \cite{EfrosNatureMat08}. The prevention of the NC ionization could explain the blinking suppression; however, it does not explain the suppression of non-radiative Auger recombination of charged excitons, biexcitons \cite{MahlerNM08,ParkPRL11, BrovelliNature11,GallandNatComm2012} and multi-excitons \cite{HtoonNL10,Garcia-SantamariaNL11}.

 The alloying  of the CdSe/CdS interface in giant CdSe/CdS NCs  could be a reason  for Auger recombination suppression, as was suggested in several papers of the Los Alamos group \cite{Garcia-SantamariaNL09, HtoonNL10,ParkPRL11,Garcia-SantamariaNL11}. This explanation relies on a theoretical prediction that softening of the carrier confined potential could reduce the Auger recombination rate by three orders of magnitude \cite{WangNature09,GraggNL10}.  Indeed, growth of the thick shell in the giant CdSe/CdS core/shell NCs  lasted for a week \cite{Garcia-SantamariaNL09, HtoonNL10,ParkPRL11,Garcia-SantamariaNL11}. The  slow growth of the shell at temperatures above room temperature suggests that the strong interface diffusion  takes place   during the NC growth. The diffusion should  lead to alloying  adjacent to the surface area and softening of the confinement potential.  This alloying assumption was confirmed recently  by the demonstration of mixed phonon replicas  in the fluorescence line narrowing (FLN) of  CdSe/CdS NCs with a thick shell \cite{Garcia-SantamariaNL11}.

This argument, however, cannot be used to  explain  the suppression of  the Auger recombination  observed by Dubertret group \cite{MahlerNM08,SpinicelliPRL09}.   The shell of their NCs is grown very fast and  that practically excludes  any inter-diffusion. In contrast, the fast growth of the lattice mismatched CdSe/CdS  core/shell NCs should obviously results in some intrinsic strains. Indeed, the samples exhibit 
a phonon spectrum of a pure CdSe NC, the frequencies of which are shifted from bulk CdSe frequencies due to the compressive strain within the NC \cite{Strain}.

 An understanding of the unusual and potentially useful properties of the giant CdSe/CdS core/shell NCs requires a theoretical description of the  fine structure of the band edge excitons, and trions.  Up to now, the energy spectra of these NC quantum dots were studied only within simplified parabolic band models \cite{Garcia-SantamariaNL11, BrovelliNatureComm11,RainoACSNano11}, which do not take into account the  complex structure of the valence band and the complicated character of inter-particle Coulomb interactions.   The theoretical description of the energy spectra of the exciton and exciton complexes in these NCs also has some objective complications, because  CdSe/CdS  hetero-structures are not well characterized.  We know that holes are always localized  in CdSe, and  that the valence band offset is no less than 500\,meV.   However, we don't know the sign of the band-offset of the conduction band nor its magnitude.   A small value of the CdSe/CdS conduction band offset suggests its complex behavior, because the band offset  sign  and magnitude should be sensitive to the strain and temperature \cite{RainoACSNano11}.

These qualitative data on the  band offset suggest that an exciton created in CdSe/CdS NCs with a thick CdS shell should look like a donor center, where a hole is strongly localized in the CdSe core while an electron  is attracted to this charged core by the Coulomb  potential. If CdSe/CdS  is the type II hetero-structure, the core potential in the NC pushes the electron into the CdS shell. In the opposite case, if CdSe/CdS is the type I heterostucture,the core creates  an additional well, which increases the  electron localization around the hole.  The temperature dependence of the CdSe and CdS energy gaps \cite{LB} suggests that 
a decrease of temperature should favor type I.
As a result, a decrease of the temperature increases the exciton binding energy and, consequently, shortens the radiative decay time of excitons, which is controlled by the electron-hole wave function overlap, and  increases the dark-bright exciton splitting.\cite{RainoACSNano11}

Due to the very strong confinement of holes, the positively charged trion which was observed  in the CdSe/CdS NCs \cite{BrovelliNature11} appears as a doubly-charged donor.  The larger Coulomb attraction of the electron to the core should significantly increase the electron binding energy and shorten the radiative decay time.  The negatively charged exciton should look like a negatively charged donor center D$^-$.  The binding energy of the second electron in the D$^-$ center is usually quite small on the order of $\sim 0.055$  of the donor binding energy. This energy  will be higher, however, in red{the} type I  CdSe/CdS core/shell structure, due to an additional attractive potential (created by the CdSe core) for an electron.   Finally, a biexciton in these structures looks like the helium atom, with two holes  localized in the core and two electrons with opposite spins occupying the same orbit.

 In this paper,  we have calculated the binding energy  and the fine structure of the exciton in CdSe/CdS NCs with a very thick shell as a function of the CdSe/CdS conduction band offset and the core radius. The calculations takes into account the complex structure of the CdSe valence band and inter-particle Coulomb  and exchange interaction.  We have also developed a general theory describing the  fine structures and radiative decay time of positively and negatively charged trions.  Using this theory  we have calculated  optical spectra of trions in giant CdSe/CdS core/shell NCs as a function of conduction-band offset and core radius.

This paper is organized as follows: In Section II and III we describe the energy spectra and wave functions of one  and two holes localized in the CdSe core. In Section IV we discuss the energy  and the wave function of an  electron in a central symmetrical potential created by localized holes and the conduction band offset. In Section V we derive the wave functions and  the energy of  excitons and positively charged trions.  
In Section VI we discuss the fine structure of these excitons and trions. In Section VII we calculate the binding energy of  the negatively-charged trions. In Section VIII we consider the radiative decay of all exciton complexes.  Finally, we discuss obtained results and compare them with available experimental data in Section IX.

\section{Hole energy spectra and wave functions}

In CdSe/CdS core shell  nanocrystals, the valence band offset is substantially large  and the holes are strongly confined in the CdSe core. For the heavy ground state holes this potential can  be considered as infinitely high in the first approximation.
The first quantum-size level of holes in a spherical NC of a semiconductor with the degenerate $\Gamma_8$ valence subband is a 
$1S_{3/2}$ state.\cite{EkimovJOSA92} This state has total angular momentum $\bm j=3/2$ and it is fourfold degenerate with respect to its projection $M=3/2,1/2, -1/2, -3/2$ on the $z$ axis. \cite{EkimovJOSA92}. The wave  functions of these four states can be written as\cite{Gelmont71}
\begin{equation}
\Psi_{M}^{h} = 2 \sum_{l=0,2} (-1)^{M-3/2} R_l(r) \sum_{m+\mu = M}
\left(
\begin{array}{ccc}
l & 3/2&3/2 \\ m&\mu&-M
\end{array}
\right) Y_{l,m} u_\mu^v \, , \label{holewf}
\end{equation}
where   $\left(_{m~n~p}^{i~~k~~l}\right)$ are Wigner 3j-symbols,
$Y_{l,m}(\theta,\phi)$ are spherical harmonics defined  in
Ref.\onlinecite{Edmans},  and $u_\mu^v $ ($\mu = \pm 1/2, \pm 3/2$) are
the Bloch functions of the fourfold degenerate valence band
$\Gamma_8$  \cite{BirPikus}. The radial functions  $R_0(r)$ and
$R_2(r)$ corresponding to an  impenetrable barrier at the CdSe/CdS interface with
radius $a$ can be written in explicit form following  Refs.
\onlinecite{EfrosPRB92} and \onlinecite{EfrosPRB96}:
\begin{eqnarray}
R_2(r)& =& {A\over a^{3/2}}\left[ j_2(\varphi r/ a)+
{j_0(\varphi )\over
j_0(\varphi\sqrt{\beta})}j_2(\varphi\sqrt{\beta}r/ a)
\right] ~,\nonumber \\
R_0(r)& =& {A\over a^{3/2}}\left[ j_0(\varphi r/a)-
{j_0(\varphi )\over j_0(\varphi\sqrt{\beta})}j_0(\varphi\sqrt{\beta}r/a)
\right]~,
\label{Rfunction}
\end{eqnarray}
  where  $j_0$ and $j_2$ are spherical Bessel functions, the constant A is determined by the normalization condition
$\int_0^a dr r^2[R_0^2(r)+R_2^2(r)]=1$ and $\beta=m_{l}/m_{h}$ is the ratio of light to heavy  hole effective masses: $m_l$ and $m_h$, respectively. The dimensionless parameter $\varphi$ is a first root of the equation  $j_0(\varphi )j_2(\sqrt{\beta}\varphi )+j_2(\varphi)j_0(\sqrt{\beta}\varphi )=0$, and describes the energy of 
the lowest hole level \cite{EfrosPRB96}: $E_{1S_{3/2}}= \hbar^2\varphi^2/(2m_ha^2)$.

The assymetry effects, which  include the intrinsic asymmetry of the hexagonal lattice structure of the crystal's field, \cite{EfrosPRB92} and the nonspherical shape of the NC  \cite{EfrosPRB93}, split the  four-fold degenerate hole state into two two-fold degenerate states with $|j_z|=3/2$ and $|j_z|=1/2$, respectively:
\be
E_{1S_{3/2},~j_z} = E_{1S_{3/2}} -  \frac{\Delta(\beta, a)}{2} (j_z^2 - 5/4) \, , \\
\label{splitting}
\ee
where $\Delta(\beta, a)= \Delta_{\rm int}(\beta)+\Delta_{\rm sh}(\beta, a)$
where $\Delta_{\rm int}$ and $\Delta_{\rm sh}$ are the intrinsic and size dependent shape contributions
which can be found, for example, in Ref. \onlinecite{biex2010}.

\section{Two hole energy spectra and wave functions}

Before considering the positively charged excitons, which can be observed in NCs with an extra hole, let us consider the energy spectra of two holes localized in the CdSe core. The application  of the Pauli exclusion principle to the two  holes  occupying the $1S_{3/2}$ level is nontrivial. Generally, the two holes with momentum $|\bm j_1| =|\bm j_2|=3/2$  according to the momentum summation rule ${\bm J}= {\bm j}_1 + {\bm j}_2$ could form four states with the total angular momentum $J=3,~2,~1~{\rm and}~0$.  The wave functions of these states can generally be written as
\begin{equation}
\Phi_{J,J_z}^{2h}({\bm r}_1, {\bm r}_2) =  (-1)^{J_z} \sqrt{2J+1} \sum_{M_1+M_2=J_z}
\left(
\begin{array}{ccc}
3/2 & 3/2& J \\ M_1 & M_2 &-J_z
\end{array}
\right)
\Psi_{M_1}^h({\bm r_1})\Psi_{M_2}^h({\bm r_2}) \, ,
\label{twoholes}
\end{equation}
where $J_z$ is the projection of the total momentum ${\bm J}$ on the $z$ axis. The Pauli hole permutation requirement applied  to the wave function described by Eq. \eqref{twoholes}, however,
allows only the two nontrivial solutions. As a result, the two holes  occupying  the $1S_{3/2}$ level  can only be in a 5-fold degenerate state with total momentum $J=2$ and a state with $J=0$.\cite{Efros-SolidStComm89,biex2010}

Hole-hole exchange interaction splits the ground biexciton into the two states with total momentum  $J=2$ and $J=0$. The straightforward calculation of the $h-h$ Coulomb interaction with the functions from Eq. \eqref{twoholes} gives the energy of the corresponding levels:
\begin{equation}
E_J(\beta)= E_{2h}(\beta) + \Delta_{\rm exch}(\beta) \left( \frac{5}{4} - \frac{J(J+1)}{6} \right)
\end{equation}
where  $E_{2h}(\beta)=2E_{1S_{3/2}}+E_{\rm Col}^{2h}$ is the
band-edge energy of the two-hole state taking into account the direct
Coulomb repulsion only,  $\Delta_{\rm exch}(\beta) = E_0(\beta) - E_2(\beta)$ is the exchange splitting
and can be written in the
following form \cite{Efros-SolidStComm89,biex2010}
\begin{equation}
\Delta_{\rm exch}(\beta) = \frac{e^2}{\kappa} \cdot \frac{32}{25} \int \int r_1^2 \, r_2^2 \, dr_1 \, dr_2 \,
\frac{r_<^2}{r_>^3} \, R_0(r_1) R_2 (r_1)  R_0(r_2) R_2(r_2) \, ,
\end{equation}
 where $\kappa$ is the dielectric constant of the semiconductor and $r_< = \ min \{r_1, r_2 \}$, $r_> = \ max \{r_1, r_2\}$. The direct Coulomb interaction energy is given by
\begin{equation}
E_{\rm Coul}^{2h}(\beta) = \frac{e^2}{ \kappa} \cdot  \int \int
r_1^2 \, r_2^2 \, dr_1 \, dr_2 \, \frac{1}{r_>} \, (R_0^2(r_1)+
R_2^2 (r_1))(  R_0^2(r_2) +R_2^2(r_2))  - \frac{5}{8}\Delta_{\rm
exc} \, .
\end{equation}
 In the spherical NCs, the
ground two hole state has total momentum
$J=2$.\cite{Efros-SolidStComm89} In the case of the
CdSe/CdS structure with impenetrable interface, when the wave
functions are described by Eq. \eqref{Rfunction}, the exchange
splitting and Coulomb energy  can be written as
\begin{eqnarray}
\Delta_{\rm exch}(\beta) = \frac{e^2}{\kappa a} \gamma(\beta) \, ,
\quad E_{\rm Coul}^{2h}(\beta) = \frac{e^2}{\kappa a} \chi(\beta)
\, .
\end{eqnarray}
In CdSe NCs, where $\beta\approx 0.28$, $\gamma(0.28) \approx 0.033$ \cite{biex2010} and $\chi(0.28) 
\approx 1.92$.

The NC asymmetry, which lifts the degeneracy of the $1S_{3/2}$ hole state in Eq.\eqref{splitting}, causes further splitting of the two hole states. The corresponding perturbation can be written as\cite{biex2010}
\begin{equation}
\hat H_{\rm as}^{2h} = \hat H_{\rm as}^{1h} (\bm r_1) + \hat H_{\rm as}^{1h} (\bm r_2) = - \frac{\Delta(\beta, a)}{2} (\hat j_{1z}^2 - 5/4) - \frac{\Delta(\beta, a)}{2} (\hat j_{2z}^2 - 5/4)\,,
\label{biexcass}
\end{equation}
where $\hat j_{1z}$ and $\hat j_{2z}$ are the operators of hole momentum projection acting on the first and second hole, respectively. The resulting two-hole states are described in Ref. \onlinecite{biex2010}.
The ground two-hole state has angular momentum projection of $J_z=0$ and energy
\begin{eqnarray}
E_0^{-}=E_{2}(\beta,a) + \frac{\Delta_{\rm exch}(\beta,a)}{2}  - \sqrt{\left(\frac{\Delta_{\rm exch}(\beta,a)}{2}\right)^2 +
\Delta(\beta,a)^2 } \, ,
\end{eqnarray}
The first excited state has $J_z=\pm 1,~{\rm and}~\pm 2$ and energy $E_2$, which is not affected by perturbations connected with NC asymmetry. Finally, the upper state has $J_z=0$ and the energy
\begin{eqnarray}
E_0^{+}=E_{0}(\beta,a) - \frac{\Delta_{\rm exch}(\beta,a)}{2}  + \sqrt{\left(\frac{\Delta_{\rm exch}(\beta,a)}{2}\right)^2 +
\Delta(\beta,a)^2 } \, ,
\end{eqnarray}
Two-hole wave functions $\Phi_0^{2h,\pm}$ (for the
energies $E_0^\pm$) and $\Phi^{2h}_{\pm 1, \pm 2}$ (for the fourfold
degenerate intermediate level $E_2$) are given
by\cite{biex2010}
 \bea \Phi_0^{2h,\pm}( {\bm r}_{h1},
{\bm r}_{h2}) &=&
\frac{1}{2} \left[ (B_2^{\pm}+B_0^{\pm}) \left( \Psi_{3/2}^h({\bm r}_{h1})\Psi_{-3/2}^h({\bm r}_{h2})-\Psi_{-3/2}^h({\bm r}_{h1})\Psi_{3/2}^h({\bm r}_{h2}) \right) + \right. \nonumber \\
&&\left.
(B_2^{\pm}-B_0^{\pm}) \left( \Psi_{1/2}^h({\bm r}_{h1})\Psi_{-1/2}^h({\bm r}_{h2})-\Psi_{-1/2}^h({\bm r}_{h1})\Psi_{1/2}^h({\bm r}_{h2}) \right) \right] \, , \nonumber \\\Phi_{\pm1}^{2h}( {\bm r}_{h1}, {\bm r}_{h2})&=& \pm \frac{1}{\sqrt{2}} \left( \Psi_{\pm 3/2}^h({\bm r}_{h1})\Psi_{\mp 1/2}^h({\bm r}_{h2})- \Psi_{\mp 1/2}^h({\bm r}_{h1})\Psi_{\pm 3/2}^h({\bm r}_{h2}) \right) \, ,\nonumber\\
\Phi_{\pm2}^{2h}( {\bm r}_{h1}, {\bm r}_{h2})&=&\pm
\frac{1}{\sqrt{2}} \left( \Psi_{\pm 3/2}^h({\bm r}_{h1})\Psi_{\pm
1/2}^h({\bm r}_{h2})- \Psi_{\pm 1/2}^h({\bm r}_{h1})\Psi_{\pm
3/2}^h({\bm r}_{h2}) \right) \, , \label{biexfunction2} \eea 
 where
\be B_0^{\pm} = \mp \, \frac{2\Delta}{\sqrt{4\Delta^2 +
D_{\pm}^2}} \, , \quad
B_2^{\pm}=\sqrt{1-(B_0^{\pm})^2}~,\label{Bs} \ee
 with
$D_{\pm}=\Delta_{\rm exch} \mp \sqrt{\Delta_{\rm exch}^2 +
4\Delta^2}$.

\section{Electron energy spectra   and wave functions}

In Section II above we have considered  holes in the
CdSe/CdS core/shell NCs assuming that they are  very strongly
confined in the CdSe core. In that case their wave functions are not
modified by the  Coulomb  interaction with the electron or by
the hole-hole interaction, and are solely determined by the radius
of the CdSe core. On the contrary, the conduction band offset in
such structures is known to be small and even the sign of the band
offset is not confirmed. As a result,  the band offset itself does not localize (or only weakly localizes) an  electron  in the CdSe core. The attractive Coulomb  potential created by the holes  plays an important role in  electron localization at the center of the core/shell NC structures. 

We will show in the next section  that the attractive  Coulomb potential created by holes localized in {\it a spherically symmetrical} CdSe core does  not have spherical symmetry.  In the first approximation, however, the ground electron state is described   by the wave function of $S$ symmetry, which  can be written as
\be
  \Psi_{\pm 1/2}^e({\bm r})=R_e^{Z}(r)Y_{00}(\Theta)u^c_{\pm 1/2}  ,
 \label{electronwf}
  \ee   
where $u^c_{S_z}=|S,S_z>$ are the Bloch wave functions of the conduction band and $S_z = \pm 1/2$ is the electron spin projection. The radial wave function $R_e^Z(r)$ is described by the radial
Hamiltonian
\begin{eqnarray}
\hat{H}_r^Z(r)= -\frac{\hbar^2}{2m_e(r)} \displaystyle{1\over r^2}
\displaystyle{d\over dr} \left( r^2\displaystyle{d\over dr}\right)
 + V^Z(r) \, ,
\label{elham} \\
m_e(r) = m_{CdSe}, \quad  V^Z(r) = V_{\rm Coul}^Z(r) + U_{\rm 0ff}, \quad
r<a \,  , \nonumber \\
m_e(r) = m_{CdS}, \quad  V^Z(r) = V_{\rm Coul}^Z(r) \, .   \quad
r>a \, \nonumber
\end{eqnarray}
 Here $U_{\rm 0ff}$ is the conduction band offset (CBO) ($U_{\rm 0ff}>0$ in type II structures and $U_{\rm Off}<0$ in type I structures), $m_{CdSe}$ and $m_{CdS}$ are the electron effective masses in CdSe and CdS, correspondingly,  and $V_{\rm Coul}^Z(r)$  is the  effective Coulomb potential of
one ($Z=1$) or two ($Z=2$) holes acting on an electron.  We use
$Z=0$ to describe the "free" electron problem in the case of the
neutral core. 

The radial wave function $R_e^Z(r)$ is completely defined  by Eq. \eqref{elham} and the standard boundary conditions at the CdSe/CdS
interface $r=a$:
\begin{eqnarray}
R_e^Z(r) |_{r<a} =R_e^Z(r) |_{r>a}, \, \nonumber \\ \frac{1}{m_{CdSe}} \frac{d
R_e^Z(r)}{dr}|_{r<a} = \frac{1}{m_{CdS}} \frac{d
R_e^Z(r)}{dr}|_{r>a} \, . \label{bc}
\end{eqnarray}

Later we will need the radial function of the "free" resident electron, which remains in NCs after negative trion recombination.  Its wave function $R_e^0(r)$ can be written as
\begin{eqnarray}
R_e^0(r) = A \left\{
\begin{array}{ll}
 \displaystyle{h_0^1\left( i \kappa a\right)\over j_0\left( k a\right)} j_0\left( k r\right) & r \le a\\ 
 h_0^1\left( i \kappa r\right) & r > a
\end{array}	
	\right.
\label{FreeWF}
\end{eqnarray}
where $A$ is the normalization constant, $h_0^1$ is the spherical Hankel function,\cite{AbramowitzStegun}
$\kappa = \sqrt{-2 m_{\mathrm{CdSe}}E_e^0}/\hbar$, 
$k = \sqrt{2 m_{\mathrm{CdS}}\left(E_e^0  + U_{\mathrm{Off}} \right)}/\hbar$ 
and $E_e^0$ is the electron ground state energy level calculated from the bottom of the
CdS conduction band. This energy is described by
the smallest solution of the following equation:
\begin{equation}
1 - ka \cot\left( ka\right) = \mu \left(1 + \kappa a \right) 
\end{equation}
where $\mu = m_{\mathrm{CdSe}}/ m_{\mathrm{CdS}} $.
It is important to note that even if CdSe/CdS core/thick shell NCs are Type I structures, the resident electron is not localized in the CdSe core with small radius $a$ and is spread over the entire CdS shell.  
The electron localization occurs only if the power of a quantum well $w_0= a\sqrt{2m_{\rm CdSe}U_{\rm Off}}/\hbar$ is larger than $w_0^{\rm cr}$, which is a solution of the following equation: 
$\cot(w_0^{\rm cr})=(1-\mu)/w_0^{\rm cr}$. For $\mu=1$ 
the electron delocalization occurs when $w_0^{\rm cr}=\pi/2$, the condition which is identical to the standard expression for the critical depth of the 3D potential allowing the electron localization, $U_{\rm Off}= \pi^2\hbar^2/(8m_{\rm CdSe}a^2)$.\cite{LL} However, the large electron effective mass in CdS resulting in $\mu =0.634$ allows  the localization in the CdSe core with 
$w_0$ smaller than $\pi/2$: $\pi/2 > w_0>w_0^{\rm cr} =1.296$.

\section{Donor-like exciton  and positively charged trion}

The electron wave function is strongly modified by a long range Coulomb potential  created by a hole or holes in the donor-like exciton or the positively charged trion in CdSe/CdS NCs with a small CBO.   Electrons are bound to the CdSe core by a
strong Coulomb potential of a single hole or two holes, $Z=1,2$
correspondingly, which are strongly confined by the core
potential, similar to an electron in singly or doubly charged
donors. This potential always  leads to some binding of electrons to the CdSe core, even in type II CdSe/CdS core/shell structures.  The attraction potential of holes is described by the
distribution of the hole positive charge within the core of the
radius $a$.

\subsection{Hole charge distribution and Coulomb potential}
The straightforward use of the hole wave function definition in Eq. \eqref{holewf} gives the charge distribution of the single hole state with $M=\pm3/2$ and $M=\pm1/2$ correspondingly:
\bea
\rho_{\pm3/2}(\bm r)=|\Psi^h_{\pm3/2}(\bm r)|^2&=&\left(R_0(r)Y_{00}+{1\over \sqrt{5}}R_2(r)Y_{20}\right)^2+{2\over 5}R_2^2(r)\left(|Y_{2\pm1}|^2+|Y_{2\pm2}|^2\right)~,~\nonumber\\
\rho_{\pm1/2}(\bm r)=|\Psi^h_{\pm1/2}(\bm r)|^2&=&\left(R_0(r)Y_{00}-{1\over \sqrt{5}}R_2(r)Y_{20}\right)^2+{2\over 5}R_2^2(r)\left(|Y_{2\mp1}|^2+|Y_{2\mp2}|^2\right)~,~
\eea
Using the definition of spherical harmonics, $Y_{l,m}$, from \cite{Edmans} we
arrive to
\be
\rho_{M}(\bm r)={1\over 4\pi}\left[R_0^2(r)+2(|M|-1)R_0(r)R_2(r)(1-3\cos^2\theta)+R_2^2(r)\right]~.
\label{eq:13}
\ee

The charge distribution of two {\it isotropic}  holes with $\Delta=0$, localized in CdSe core,
$\rho_{J,J_{z}}^{2h}(\bm r)$ can be obtained  from the  two-hole
wave functions $\Phi_{J,J_{z}}^{2h}$ defined in Eq.
\eqref{twoholes}: \be \rho_{J,J_{z}}^{2h}(\bm r)=\int d^3r_{h1}
d^3r_{h2}|\Phi_{J,J_{z}}^{2h}( {\bm r}_{h1}, {\bm
r}_{h2})|^2[\delta(\bm r-\bm r_{h1})+\delta(\bm r-\bm r_{h2})] \ee
To obtain the hole charge distribution for $\Delta\neq 0$ one has to use the two hole wave functions described  in  Eq. (\ref{biexfunction2}). The calculations show that  for the four-time degenerate hole state $E_2$, which
is characterized by the angular momentum projections $J_z=\pm 1,
\pm 2$, the charge distribution is spherical and is described as
\be \rho^{2h}_{\pm1, \pm 2}(r)={2\over
4\pi}\left[R_0^2(r)+R_2^2(r)\right] \label{eq:17} \ee For the
upper, "$E^+_0$",  and lower, "$E^-_0$", two hole states with
$J_z= 0$ similar calculations result
 \be \rho^{2h,\pm}_{0}(\bm
r)={2\over 4\pi}\left[R_0^2(r)+R_2^2(r)+2B_0^\pm B_2^\pm
R_0(r)R_2(r)(1-3\cos^2\theta) \right] \label{eq:18} \ee
In the limit  $\Delta\gg \Delta_{exch}$ Eq. \eqref{eq:18} is simplified
significantly:
\be \rho^{2h,\pm}_{0}(r)={2\over
4\pi}\left[R_0^2(r)+R_2^2(r)\mp R_0(r)R_2(r)(1-3\cos^2\theta)
\right]~. \label{eq:19} \ee
Comparison of Eq. \eqref{eq:19} with
Eq. \eqref{eq:13} shows clearly that in the case $\Delta\gg
\Delta_{exch}$,  the lowest level is created from two heavy holes,
while the uppermost level is created by two light holes. Similar
comparison  of Eq. \eqref{eq:17} with Eq. \eqref{eq:13} shows that
the middle level is always created by one light and one heavy
hole.

One can show that only the spherical part of the hole charge distribution contributes to the  ground $S$ electron state in first order perturbation theory. For the single ($Z=1$) and double ($Z=2$) hole states this distribution can be written as
\be
\rho^{Z}_{0}(r)={Z\over 4\pi}\left[R_0^2(r)+R_2^2(r)\right]
\ee

The Coulomb potential created by  the hole spherical charge
distribution  can be written as
 \be V^Z_{\rm Coul}(r)=-{Ze^2\over
\kappa}\int_0^adr'^3{\rho^{Z}_{0}(r')\over |\bm r -\bm r'|}
\label{eq:21} \ee
 where $\kappa$ is the dielectric constant, which
is considered to be the same in   both CdSe and CdS.   Equation
\eqref{eq:21} can be rewritten as:
 \be V^Z_{\rm Coul}(r)=-{Ze^2 4\pi\over
\kappa }\left[{1\over r}\int_0^r
dr'r'^2\rho^{Z}_{0}(r')+\int_r^adr'r'\rho^{Z}_{0}(r')\right]~.
\label{potentialV} \ee
 This potential outside the CdSe core, $r>a$, always has a standard Coulomb 
 orm, $V(r)= Ze^2/(\kappa r)$, due to the wave function normalization 
 condition,  $\int_0^a drr^2 [R_0^2(r)+R_2^2(r)]=1$.  Inside the core $V'(r)|_{r=0}=0$, and  the potential depth   $V(0)=-(Ze^2 4\pi/ \kappa) \int_r^adrr\rho^{Z}_{0}(r)$ depends only on the ratio $\beta$ of the light- to heavy-hole effective masses. In CdSe where $\beta=0.28$; $V(0)= 2.684Ze^2/(\kappa a)$.

\subsection{Electron wave function and binding energy}

We will consider only CdSe/CdS core/shell NCs with a very thick shell because  these NCs show 
the most interesting and useful properties.
 From now on all calculations will be conducted for heterostructure with unlimited CdS shell thickness;  later we will discuss the limitations of our  approach.

The ground state of a weakly bound electron in the donor-like exciton or positively charged trion created in CdSe/CdS  is described by the Hamiltonian of Eq. \eqref{elham} with Coulomb potential, $V_{\rm Coul}^Z$,  described by Eq. \eqref{potentialV}.   
Introducing dimensionless distance from the NC center, $x=r/a$, and dimensionless energies $\bar{E}= E/(\hbar^2/2m_{\rm CdSe} a^2)$, we obtain the dimensionless Hamiltonian,
\begin{eqnarray}
\hat{H}_r^Z(x)=\left\{\begin{array}{lr}
-\displaystyle{1\over x^2} \displaystyle{d\over dx}
\left( x^2\displaystyle{d\over dx}\right)
 + \displaystyle{Z\over \eta_0} \bar{V}\left(x \right)+\bar{U}_{\rm off}~&,~x<1 \\
  -\mu \displaystyle{1\over x^2} \displaystyle{d\over dx}\left( x^2{d \over dx}\right)
 + {Z\over \eta_0} \bar{V}\left(x \right) ~&,~x>1\end{array}\right.
\label{Dimensionless}
\end{eqnarray}
for the radial component of the electron  wave function, $R^Z_e(x)$. In Hamiltonian 
$\hat{H}_r(x)$ we use  
$\bar{V}(x)=2V_{\rm Coul}(r)/(Ze^2/\kappa a)$, $\bar{U}_{\rm Off}=U_{\rm Off}/(\hbar^2/2m_{\rm CdSe} a^2)$.
All energies are calculated from the bottom of the CdS conduction band,
 and the confinement potential $U_{\rm Off}$ created by the conduction band offset is negative in 
 type I, and positive in type II, CdSe/CdS hetero-structures.
We also introduce the unitless parameter $\eta_0=a_B/a$ that characterizes the
strength of Coulomb potential created by holes,  where $a_B =\hbar^2 \kappa/ \left(m_{\rm CdSe} e^2\right)$ is the Bohr radius.

A central parabolic part of the hole potential described by Eq. \eqref{potentialV}
becomes important if $Z/\eta_0$ is large.  The wave function of the 1S ground state should be close to the wave function of the harmonic oscillator. In the opposite case of small $Z/\eta_0$, the core potential is weak and the electron moves  mainly in the Coulomb-like potential at large
distances from the core. To describe both above-mentioned limits, we select  a trial radial function in the  following form:
\begin{eqnarray}
R_e^Z(\alpha,\zeta;x) = \left\{
\begin{array}{ll}
(Ax^2+B) e^{\left(1 - x^2 \right)/\alpha^2 } & x \le 1\\
D e^{\left( 1 - x \right)/\zeta } & x > 1
\end{array}
\right.
\label{Trial1}
\end{eqnarray}
where $\alpha$ and $\zeta$ are dimensionless variation parameters, which will be found using the variational principle,
and constants $A$,$B$, and $D$, which are completely determined by the interface boundary and normalization conditions. The standard boundary conditions at the CdSe/CdS interface ($x = 1$) described by Eq. \eqref{bc} gives
\begin{eqnarray}
R_e^Z(\alpha,\zeta;x) = \left\{
\begin{array}{ll}
B e^{\left(1-x^2\right)/\alpha ^2} \left[  1 + x^2
\displaystyle{ \left(2\zeta - \alpha^2 \mu \right)\over 2 \left(\alpha ^2 -1\right) \zeta +\alpha ^2 \mu} \right] & x \le 1\\
B e^{ \left(1-x \right)/\zeta}
\left.\displaystyle{ 2 \alpha^2 \zeta \over 2 \left(\alpha ^2 -1\right) \zeta +\alpha ^2 \mu}\right. & x > 1 ~,
\end{array}
\right.
\label{Trial1ABCD}
\end{eqnarray}
where $B$ can be found from the normalization condition $\int_0^\infty dx x^2 R_e^2(\alpha,\zeta;x)=1$ 
(see Appendix).

Using Eq. \eqref{Dimensionless}
we can find the expectation value for the energy, $\bar{E}^Z$, of one electron:
\be
\langle \bar{E}^Z\rangle_e = \int\limits_{0}^{\infty}
R_e^Z(x)\hat{H}_r^Z(x)  R_e^Z(x)
 x^2 dx
\label{Trail1Total}
\ee

\begin{figure}[htb]
\vskip-0.1truecm
\begin{center}
\epsfig{file=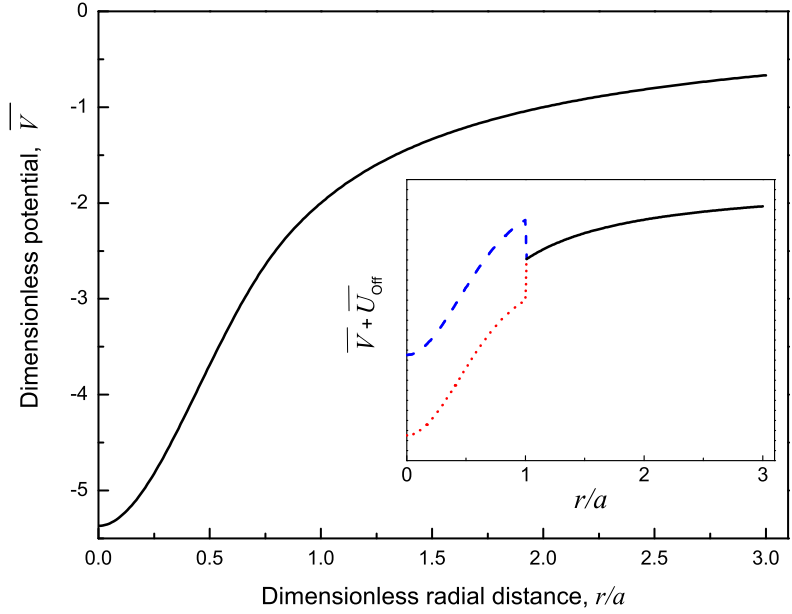, angle=0, width=0.75\textwidth}
\end{center}
\vskip-0.7truecm
\caption{Dimensionless Coulomb potential $\bar{V}\left( x\right)$ created by  one hole ($Z = 1$). Insert shows schematically the total potential created by the sum of the Coulomb potential acting on the electron and the conduction band offset.}
\label{potential}
\end{figure}

To simplify the calculation we approximate the potential, $\bar{V}(x)$ which is inside the core ($x<1$) by the polynomial $\bar{V}(x)\approx p_0+ p_2 x^2+ p_3 x^3+ p_4 x^4+p_5 x^5$, where $p_0=-5.368$ 
and $p_1=0$ were selected to satisfy the exact potential properties at $x=0$.  
Outside the core ($x > 1$) the potential is given by $\bar{V}=-2/x$ 
according to Eq.(\ref{potentialV}).  The potential and the potential derivatives should be continuous
 at $x=1$, which allows us to determine $p_4 = 14.84257478 - 3 p_2 - 2 p_3$ and $p_5 = -11.4740598  + 2 p_2 + p_3$.  Next, we vary the coefficients $p_2$ and $p_3$ to fit the
polynomial to the function $\bar{V}\left( x\right)$, the form of which depends only on the
parameter $\beta$. The best fit of $\bar{V}\left( x\right)$ for $\beta = 0.28$ was achieved using 
$p_2 = 10.6559838$ and $p_3 = -7.111613$.
Figure \ref{potential} shows the dimensionless potential $\bar{V}\left( x\right)$. 
The COB is introduced via $\bar{U}_{\rm Off}$  
One can see that when the band offset is positive, the electron is pushed away from the CdSe core. 
The negative band offset added to the Coulomb attractive potential stimulates electron localization.
The second hole localized in the CdSe core just increases
the magnitude of the attractive potential by a factor of two.
\vskip-0.1truecm

Using this approximation we were able to develop an analytical expression for the electron  binding energy
$\langle \bar{E}^Z\rangle_e (\alpha,\zeta)$,  via  parameters $\alpha$ and $\zeta$ and find 
the absolute minimum of the total energy  and a corresponding
pair of $\zeta$ and $\alpha$ for given $\mu$, $\eta_0$, $Z$, and  the band offset $U_{\rm Off}$ (see 
online Supplemental Material).

\begin{figure}[htb]
\vskip0.1truecm
\begin{center}
\epsfig{file=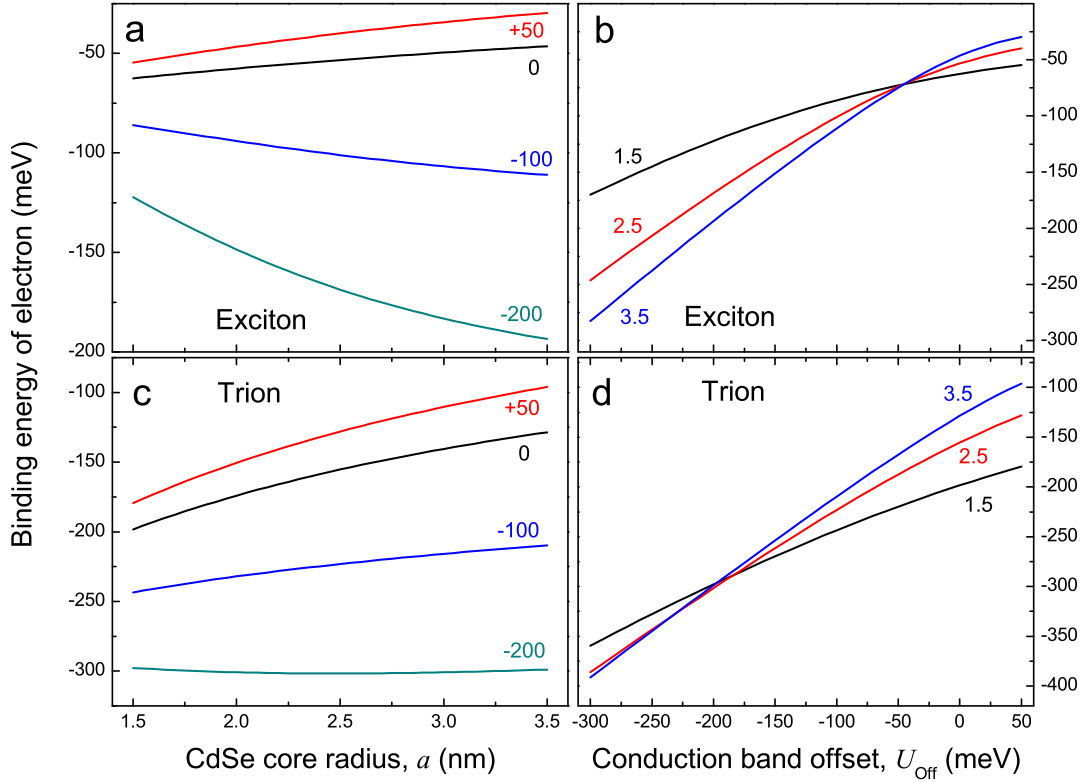, angle=0, width=1.0\textwidth}
\end{center}
\vskip-0.7truecm
\caption{Binding energy of an electron in the Coulomb potential created by one and two holes localized in 
the CdSe core calculated for various core radii and conduction band offsets.  
The dependence of the electron binding energy on the CdSe core radius, $a$, for one hole (Panel a) and two holes 
(Panel c). The dependence of the electron binding energy on the conduction band offset 
$\bar{U}_{\rm Off}$ for one (Panel b)  and two holes  (Panel d). }
\label{BindingEnergy}
\end{figure}

Figure \ref{BindingEnergy} shows the binding energy of an electron for various core radii and band offsets, calculated for the cases when the CdSe core is charged with one ($Z=1$) and two ($Z=2$) holes. The energy is calculated from the bottom of the of the CdS conduction band.
 The size  dependences are calculted for a CdSe core with a 2.5 nm radius and  shown for 
five band offsets: $u=50, 0, -50, -100, -200$\, meV.   One can see that in NCs with small negative or positive
band-offset, this energy decreases with a radius increase, while in the case 
of -100 and -200\, meV, it increases following the bottom of the conduction band. The decrease of the binding energy with radius increase in NCs  with small negative or positive band-offset  is connected with the fact that electron motion in such NCs is controlled by  Coulomb potential, which decreases with NC size.
\vskip0.1truecm
Figure \ref{BindingEnergy}  shows the  dependence of the electron binding energy on the band offset,$\bar{U}_{\rm Off}$, in NCs with core radius 1.5, 2.5, and 3.5 nm radius for the single and doubly charged CdSe core.  The band offset depends on temperature \cite{RainoACSNano11}, and the shift of the exciton line can be observed in temperature dependence.

\section{The fine structure of excitons and positively charged trions}

The fine structure of the excitons and trions  is strongly affected by the 
electron-hole exchange interaction, which can be written in spherical approximation as
\be
H_{\rm exch}= -{2\over 3}a_0^3\delta({\bm r}_e-{\bm r}_h)\varepsilon_{\rm exch}({\bm \sigma} \cdot{\bm j}) \, 
,
\label{eq:28}
\ee
where,$a_0$ is the lattice constant,  $\varepsilon_{\rm exch}$ is the strength of the exchange interaction (in energy units), ${\bm \sigma}$ are the Pauli matrices, and ${\bm j} = j_x,j_y,j_z$ are three matrices of the momentum 3/2. The  strength of the exchange interaction in cubic and hexagonal semiconductors can be determined from the bulk exciton fine structure. In CdSe this constant extracted from the bulk exciton splitting
$\varepsilon_{\rm exch}=450$  meV. \cite{EfrosPRB96}

Averaging Eq. \eqref{eq:28} over electron and hole wave functions in the basis of the 8-fold degenerate exciton state \cite{EfrosPRB96}, one can arrive at the following expression for the spin-spin exchange Hamiltonian between an electron and a hole:
\be
H_{\rm exch}^1= -\eta_1 ({\bm \sigma} \cdot{\bm j}) \, ,
\ee
 where the exchange parameter $\eta_Z$ depends strongly on the number of holes in CdSe core:
\be
\eta_Z=\frac{a_0^3}{6\pi}\varepsilon_{\rm exch}\int_0^a dr r^2[R_e^Z(r)]^2[R_0^2(r)+0.2R_2^2(r)],
\label{exchange}
\ee
 due to the strong effect of the core charge on the electron radial function, $R_{e}^Z(r)$. The radial functions $R_{e}^Z(r)$ and $R_{0,2}(r)$ are defined in Eqs.\eqref{Trial1ABCD} and \eqref{Rfunction}, respectively. Integration in
Eq. \eqref{exchange} goes only to the core radius $a$ because in our approximation the hole wave function $R_{0,2}(r)$ vanishes at the CdSe/CdS interface. One can see from Eq. \eqref{exchange}, that if an electron is only weakly localized in the CdSe  core, then   the parameter $\eta$  is small. This parameter is proportional to  
$(a_0/a\zeta)^3\ll 1$ where $a\zeta$ is a localization radius of the electron.

\begin{figure}[htb]
\vskip-0.1truecm
\begin{center}
\epsfig{file=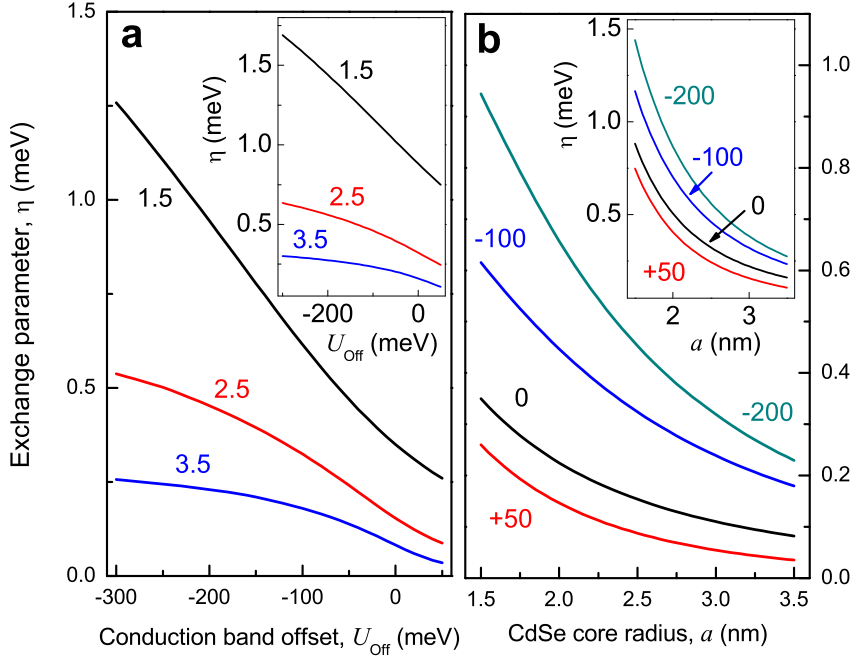, angle=0, width=0.8
\textwidth}
\end{center}
\vskip-0.7truecm
\caption{Dependence of the exciton exchange parameter $\eta_1$ on the conduction band offset, $U_{\rm Off}$ (a), and CdSe core radius, $a$ (b). Insets show the same dependences for the trion exchange parameter $\eta_2$.}
\label{Figexch}
\end{figure}

In Fig. \eqref{Figexch}a we show the dependence of the exchange energy $\eta_1$ of the  exciton as a function  of the conduction band offset for several CdSe core radii.   One can see that increase of  $U_{\rm Off}$ leads to an increase of electron localization in the core always increases $\eta_1$.   Figure \eqref{Figexch}b we show the dependence of exciton exchange energy $\eta_1$ as a function of the CdSe core radius for several NCs band offsets. For all reasonable set of radii, $\eta_1$ increases with the decrease of the NC radius.
\vskip-0.1truecm

The ground exciton state in spherical NC of zinc-blende
semiconductors is  characterized by the total exciton momentum
${\bm F}={\bm j} + {\bm S}$. For the ground 5-fold degenerate Dark exciton state $F= 2$,  and 
for the excited 3-fold degenerate Bright exciton state $F= 1$.  In
cubic semiconductor NCs with nonspherical shape and NCs with
wurzite lattice structure these states are split into 5 excitons,
each of them characterized by the projection of the total
mometum on the hexagonal axis $F_z$.\cite{EfrosPRB96} 
The respective wave functions are constructed from the hole
functions of Eq. \eqref{holewf} and electron functions of Eq.
\eqref{electronwf}, with the radial functions $R_e^{Z=1}$ for the
electron moving in the Coulomb potential $V_{\rm Coul}^{Z=1}$ created
by one hole localized in the CdSe core.

 The band edge exciton fine structure is completely characterized by two
parameters $\eta_1$ and $\Delta(a,\beta)$ defined above. In
spherical cubic NCs, the splitting  between Bright and Dark
excitons is given by $E_2-E_1=4\eta_1$.
 In the case corresponding to $\Delta(a,\beta)\gg \eta_1$, the splitting between  
 the ground Dark exciton with $F_z=\pm2$  and the
 first excited Bright exciton exciton with $F_z=\pm1$ is $3\eta_1$.
  The last case is most probably realized in presently grown CdSe/CdS NCs.\cite{BrovelliNatureComm11}
  In the more general case, corresponding to  $\Delta(a,\beta)\sim \eta_1$,  the exciton fine structure is described by expressions which can be found in Refs. \onlinecite{EfrosPRB96}, \onlinecite{biex2010}, and \onlinecite{KlimovBook}.

The operator for the electron-hole exchange interaction in the positively charged trion can be written as
\be
H_{\rm exch}({\bm r}_{1})+H_{\rm exch}({\bm r}_{2})= -{2\over 3}a_0^3\varepsilon_{\rm exch}\left[\delta({\bm r}_e-{\bm r}_1)({\bm \sigma} \cdot{\bm j}_1)+\delta({\bm r}_e-{\bm r}_2)({\bm \sigma} \cdot{\bm j}_2)\right]\,
\label{eq:31}
\ee
where ${\bm r}_1$, ${\bm r}_2$ and  ${\bm j}_1$, ${\bm j}_2$ are coordinates and spin matrices of two holes, respectively.  The averaging of Eq. \eqref{eq:31} over radial components of the two hole and electron wave functions results in the  effective electron-hole spin exchange Hamiltonian
\be
H_{\rm exch}= - 2\eta_2 ({\bm \sigma} \cdot{\bm J})\, , {\bm J}={\bm j}_1+{\bm j}_2 \, ,
\ee
  where $\eta_2$ is defined in Eq. \eqref{exchange}.  The dependences of the trion exchange  parameter $\eta_2$  on the core radius  and the conduction band offset are shown in insets of Fig. \eqref{Figexch}. One can see that due to stronger attraction of the electron to the CdSe cores coursed by the second hole, the trion's $\eta_2$ are larger than those for the exciton.

In the absence of the hole level splitting  ($\Delta(\beta, a)=0$), the states of the positively charged trion are
 characterized by the total momentum ${\bm G}={\bm j}_1+ {\bm j}_2+ 1/2{\bm \sigma}={\bm J} + {\bm S}$ and the
trion wave functions can be generally found as
\begin{equation}
\Phi_{G,g}^{tr}({\bm r}_e,{\bm r}_{h1}, {\bm r}_{h2}) =
(-1)^{g-1/2} \sqrt{2G+1} \sum_{J_z+S_z=g} \left(
\begin{array}{ccc}
J& 1/2& J \\ J_z & S_z &-g
\end{array}
\right) \Phi_{J,J_z}^{2h}({\bm r}_{h1}, {\bm
r}_{h2})\Psi^e_{S_z}({\bm r}_e) \, , \label{trionwf}
\end{equation}
where $g$ is the projection of the total momentum ${\bm G}$ on the $z$
axis, $S_z=\pm 1/2$ is the projection of the electron spin on
the $z$ axis, $\Phi_{J,J_z}^{2h}$ is the two-hole wave function described by Eq. (\ref{twoholes}),  and $\Psi^e_{S_z}$ is the electron wave function described  by Eq. (\ref{electronwf}) with the radial function
$R_e^{2}$ for the electron moving in the Coulomb potential created by two holes.
The ground trion state has total momentum $G=5/2$ and the energy
$E_{5/2}=E_2-2\eta_2$; the next one has $G=3/2$ with the energy
$E_{3/2}=E_2+3\eta_2$; and, finally, the upper state has   $G=1/2$ with the energy $E_{1/2}=E_0$. The
last trion state is not affected by electron hole exchange
interaction because the electron interaction with two different
holes exactly compensate each other.  The splitting of the two lowest
hole states $E_2$ coursed by the exchange interaction is
$E_{5/2}-E_{3/2}=5\eta_2$. The order of two excited trion states
depends on the relative strength of electron-hole or hole-hole
exchange interactions.  All these trion states are twofold degenerate with respect of
the momentum projection $g$, and their wave functions, $\Phi^{tr}_{G,g}$,  can be written
 \bea
\Phi^{tr}_{1/2,\pm 1/2}&=&\Phi^{2h}_{0,0}\Psi^e_{\pm 1/2}~, ~
\Phi^{tr}_{3/2,\pm 1/2}=\pm \sqrt{\frac{3}{5}}\Phi^{2h}_{2,\pm 1}\Psi^e_{\mp 1/2}\mp \sqrt{\frac{2}{5}}\Phi^{2h}_{2,0}\Psi^e_{\pm 1/2}~,\nonumber\\
\Phi^{tr}_{3/2,\pm 3/2}&=&\pm
\sqrt{\frac{1}{5}}\left(2\Phi^{2h}_{2,\pm 2}\Psi^e_{\mp
1/2}-\Phi^{2h}_{2,\pm 1}\Psi^e_{\pm 1/2}\right)~,~
\Phi^{tr}_{5/2,\pm 1/2}=\sqrt{\frac{2}{5}}\Phi^{2h}_{2,\pm 1}\Psi^e_{\mp 1/2}+\sqrt{\frac{3}{5}}\Phi^{2h}_{2,0}\Psi^e_{\pm 1/2}~,\nonumber\\
\Phi^{tr}_{5/2,\pm 3/2}&=&
\sqrt{\frac{1}{5}}\left(\Phi^{2h}_{2,\pm 2}\Psi^e_{\mp
1/2}+2\Phi^{2h}_{2,\pm 1}\Psi^e_{\pm 1/2}\right)~,~
\Phi^{tr}_{5/2,\pm 5/2}=\Phi^{2h}_{2,\pm 2}\Psi^e_{\pm
1/2}~,\label{tr55} \eea .

When the ground state of holes is split by a crystal field or when the NC shape is asymmetric, the related perturbation can be written  in the basis of the  trion functions  of Eq. \eqref{tr55}. The perturbation does not mix the the trion state with different signs of the spin projections.  For the positive spin projections
 ($g=5/2, 3/2$, and 1/2) the perturbation has the following form:
\begin{equation}
\left(
\begin{array}{c|cccccc}
& |1/2,1/2\rangle & |3/2,1/2\rangle & |5/2,1/2\rangle& |3/2,3/2\rangle & |5/2,3/2\rangle& |5/2, 5/2\rangle\\
\hline
|1/2,1/2\rangle &  E_0 &  \sqrt{\frac{2}{5}}\Delta  & -\sqrt{\frac{3}{5}}\Delta  & 0 & 0& 0 \\
|3/2,1/2\rangle& \sqrt{\frac{2}{5}}\Delta & E_2+3\eta_2  & 0 & 0 &0 & 0 \\
|5/2,1/2\rangle & -\sqrt{\frac{3}{5}}\Delta  &0  & E_2-2\eta_2 & 0 & 0& 0  \\
|3/2,3/2\rangle& 0 & 0 & 0 & E_2+3\eta_2 & 0& 0 \\
|5/2,3/2\rangle& 0 & 0 & 0 &0  &E_2-2\eta_2 &0  \\
|5/2,5/2\rangle & 0 & 0 & 0 & 0 &0 & E_2-2\eta_2
\end{array} \right) \, \\
\label{eq:34}
\end{equation}
A very similar matrix describes  the perturbation created by  nonzero $\Delta$ on the states with  negative angular momentum projections: $g=-5/2, -3/2$, and -1/2.  That matrix can be obtained by replacing states $|G,g'\rangle$ in Eq.\eqref{eq:34} by states with $|G,-g'\rangle$, where $g'=5/2,3/2,~{\rm and}~1/2$. The only difference is the sign of the matrix element
 taken between states $|1/2,-1/2\rangle$ and $|3/2,-1/2\rangle$, which in the last case is $-\sqrt{2/5}\Delta$.

One can see from Eq. \eqref{eq:34}, that NC asymmetry does not affect the three trion  states $|5/2,\pm 5/2\rangle$, $|5/2,\pm 3/2\rangle$ and $|3/2,\pm 3/2\rangle$. The   energies of these states are
\be E_{5/2}=E_2- 2\eta_2~,~
E_{3/2}=E_2+3\eta_2 \, ,
\ee
and their wave functions, $\Phi^{tr}_{3/2,\pm 3/2}$ for $E_{3/2}$, and $\Phi^{tr}_{5/2,\pm 3/2}$ and $\Phi^{tr}_{5/2,\pm 5/2}$ for $E_{5/2}$ are the same as in the spherical case.

The other three trion states, $E_{1/2}^\pm$ and $E_{1/2}$ can be found
as  solutions of the following third order equation: \be
(E_2-E-2\eta_2)(E_2-E+3\eta_2)(E_0-E)-\Delta^2(E_2-E+\eta_2)=0 \, ,
\label{trionlev} \ee
 In the case of small electron-hole and hole-hole exchange interactions  $\eta_2 \ll \Delta$, $\Delta_{exc} \ll \Delta$,
  the  approximate solutions of Eq. \eqref{trionlev} yield 
\be E_{1/2}^{\pm}\approx E_0^{\pm}\approx E_2 +\Delta_{exc}/2 \pm
\Delta \, , \, E_{1/2}\approx E_2+\eta_2~. 
\ee
 The wave functions of these trion states are  $\Phi^{tr, \pm}_{\pm 1/2}= \Phi^{2h,\pm}_{0}\Psi^e_{\pm 1/2}$ and $\Phi^{tr}_{\pm 1/2}= \Phi^{2h}_{2\pm 1}\Psi^e_{\mp 1/2}$, respectively.
  
In the limit of small exchange interaction, the order of the trion levels is independent of the relation between $\Delta_{exc}$ and $\eta_2$.
  The fine structure of the positively charged  trion is, ordered upward in energy, the ground  twofold degenerate $E_{1/2}^{-}$ level,  followed by  the fourfold degenerate
 $E_{5/2}$ level, followed by the twofold degenerate $E_{1/2}$ and $E_{3/2}$ levels, and  finally  by the upper twofold degenerate $E_{1/2}^{+}$ level. The lowest
and the uppermost trion states in the limit of large $\Delta$ are formed
from the lowest and uppermost two-hole states with a total zero
momentum projection of two holes $J_z=0$, while all intermediate
states are formed from the medium fourfold degenerate two-hole
level with $J_z=\pm 1, \pm 2$.  The medium  level of the trion would be eightfold degenerate if one neglected the splitting  caused by the electron-hole exchange interaction proportional to $\eta_2$. 

The limit of small exchange interaction should describe quite well the CdSe/CdS NCs which
have a core radius $\sim 20$ {\AA}. In these NCs, the hole-hole
exchange interaction,  $\Delta_{exc} \sim 5$\,meV and
$\Delta(\beta,a)\sim 20$ meV. The calculations of the electron-hole
exchange interaction in Fig. \ref{Figexch} show that $\eta_2$ is also much
smaller than $\Delta(\beta, a)$. 

\section{Negatively charged trions}

Let us consider now a negatively charged trion formed by the optically excited electron-hole pair and resident electron. 
The  total energy of the  negatively charged  trion can be generally described as  a sum of the two  total energies of electrons interacting with the hole described in our case by Eq. \eqref{Trail1Total} and the energy of the electron repulsion. In the case of the small conduction band offsets, the binding energy of the trion is mainly controlled by electron-electron and electron-hole Coulomb interactions as is true for the negatively charge hydrogen atom. The resulting binding energy in such a case is significantly  smaller than the binding energy of the exciton  because the Coulomb attraction between a second electron and the hole is almost exactly compensated by the Coulomb repulsion between electrons. 

The studies of negatively charged hydrogen atoms (see for example Ref. \onlinecite{BS}) 
demonstrate that the binding state of doubly charged atom exist only for a singlet state 
of two electrons occupying the two radial orbits of $S$ symmetry. This requirement decreases the Coulomb repulsion energy between two electrons. That is why we select the  two-electron  wave functions of the negative trion
$\Psi_{2e}({\bm r_1},{\bm r_2})$  as the
product of the  singlet spin wave function (antysimmetrical) and
symmetrical  radial function of two $S$-symmetry electrons in the following form:
 \bea
\Psi_{2e}({\bm r_1},{\bm
r_2})=\frac{1}{\sqrt{2}} (u_{1/2}^c(e1)u_{-1/2}^c(e2)-
u_{-1/2}^c(e1)u_{1/2}^c(e2))  Y_{00}(\Theta_{e1})Y_{00}(\Theta_{e2})R_{2e}(r_1,r_2)
\, ,
\eea
where $R_{2e}$ is defined as
 \be
 R_{2e}(r_1,r_2)=A_{2e}[R_e^{1}(\alpha_1,\zeta_1;r_1/a)R_e^{1}(\alpha_2,\zeta_2;r_2/a)+R_e^{1}(\alpha_1,\zeta_1;r_2/a)R_e^{1}(\alpha_2,\zeta_2;r_1/a)] 
\label{WF2e} 
\ee 
 with 
$R_e^{1}(\alpha,\zeta;r/a)$  defined in Eq. \eqref{Trial1ABCD}
and $A_{2e}$ defined by the normalization condition
$\int_0^\infty\int_0^\infty
|R_{2e}(r_1,r_2)|^2r_1^2dr_1r_2^2dr_2=1$. 

Using this approach we can write down the total dimensionless
energy  of two electrons in
the trion 
 \be
\langle\bar{E}\rangle_{2e}= 2\int_0^\infty
R_{2e}(r_1,r_2)\hat{H}_r(r_1)R_{2e}(r_1,r_2){d^3r_1\over
4\pi}{d^3r_2\over 4\pi}+ E_{\rm Coul}^{2e} \, ,  \label{eq:45} \ee where
the repulsion energy between two electrons $E_{\rm Coul}^{2e}$ can be
found as
 \bea
E_{\rm Coul}^{2e} = \frac{e^2}{\kappa}
 \int {d^3r_1\over 4\pi}{d^3r_2\over 4\pi}
{|R_{2e}(r_1,r_2)|^2\over |\bm r_1-\bm r_2|} =
\frac{e^2}{\kappa}\int
{r_1^2dr_1}{r_2^2dr_2}{R_{2e}^2(r_1,r_2)\over r_{>}} \eea

The two-electron trial wave function defined in Eq.\eqref{WF2e} is described by  
four fitting parameters,  and it generally describes the electron motion  on two radial orbits characterized by two different sets  $\alpha_1$, $\zeta_1$ and $\alpha_2$, $\zeta_2$. It is important to note that if $\alpha_1\neq \alpha_2$ and  $\zeta_1\neq \zeta_2$, the energy of two electrons in Eq. \eqref{eq:45} is not reduced to the kinetic energy and repulsion of  uncorrelated electrons occupying two different orbits. The total energy of electrons has a contribution from the interference  of both orbits and depends on all four parameters.

\begin{figure}[htb]
\vskip-0.1truecm
\begin{center}
\epsfig{file=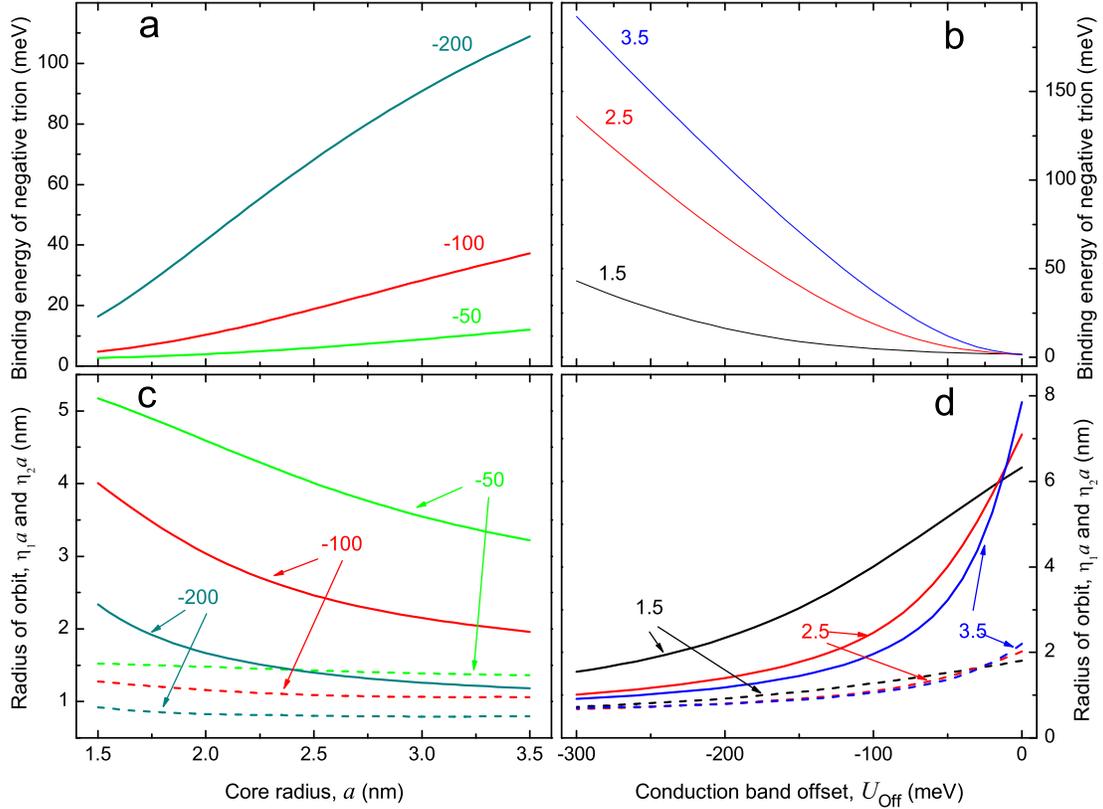, angle=0, width=1.0\textwidth}
\end{center}
\vskip-0.7truecm
\caption{Binding energy of the negatively charged trion in CdSe/CdS core thick/shell NCs. (a) - dependence of the binding on the core radius, $a$, calculated for three band offset $U_{\rm Off}$=-50, -100, and -200\,meV.  (b) - dependence of the binding on the band offset $U_{\rm Off}$ calculated for three radii $a=1.5$, 2.5, and 3.5\,nm. Panels c and d show the dependence of the orbit radii $\zeta_1 a$ and $\zeta_2a$ as functions of  the core radius, $a$,  and  of the band offset $U_{\rm Off}$, respectively.}
\label{trionBindingEnergy}
\end{figure}

 One can find the analytical expression of  the expectation energy $\langle\bar{E}\rangle_{2e}$ as a function of four parameters
  in the online SM. This expression developed within {\it Mathematica} were used to find the  binding energy of two electrons and a hole.
   The trion binding energy, which is  the difference between the total energy of the trion and the binding energy of the exciton (see Fig. \ref{BindingEnergy}),
   is shown in Fig. \ref{trionBindingEnergy} as a function of of the NC core radius and band offset.
  
\vskip-0.1truecm
    One can see that for large negative values of $U_{\rm Off}$, the CdSe/CdS NCs are well-defined type-I structures. The strong confinement allows the two electrons to occupy a ground state level with two opposite spins leading to a large trion binding energy (see Fig. \ref{Figovlp}). Decrease  of $|U_{\rm Off}|$  decreases the confinement, and the trion binding energy is mainly controlled by the electron-electron and the electron-hole Coulomb interactions. The Coulomb repulsion between two electrons forces them to occupy different orbits, which reduces the repulsion and leads to a small, but positive binding energy for the trion.  The orbits are characterized by the two radial wave functions, $R_e^1(\alpha_1,\zeta_1,r/a)$ and $R_e^1(\alpha_2,\zeta_2,r/a)$, the spatial extension of which is characterized by parameters $\zeta_1 a$ and  $\zeta_2 a$. The radius of these two orbits  is shown in Fig. \ref{trionBindingEnergy}c and \ref{trionBindingEnergy}d as a function of core radius and band offset.  One can see that one of the orbits is mainly localized in the CdSe core while the other explores the CdS shell. The radius of the large orbit obviously increases with the decrease of $|U_{\rm Off}|$ for any core radius, as one can see in Fig. \ref{trionBindingEnergy}c.  The CBO dependence shown in Fig. \ref{trionBindingEnergy}d is more complex, however. The radius of the external orbit is smaller in the NC with a large core radius at large CBOs, but it becomes larger at small CBOs. This unusual behavior is connected with the increased role that the hole Coulomb attraction potential plays in electron localization in core/shell NCs with a small CBO.

\section{Optical properties of excitons and trions}
\subsection{Optical transition probabilities of the band edge excitons}

\begin{figure}[htb]
\vskip-0.1truecm
\begin{center}
\epsfig{file=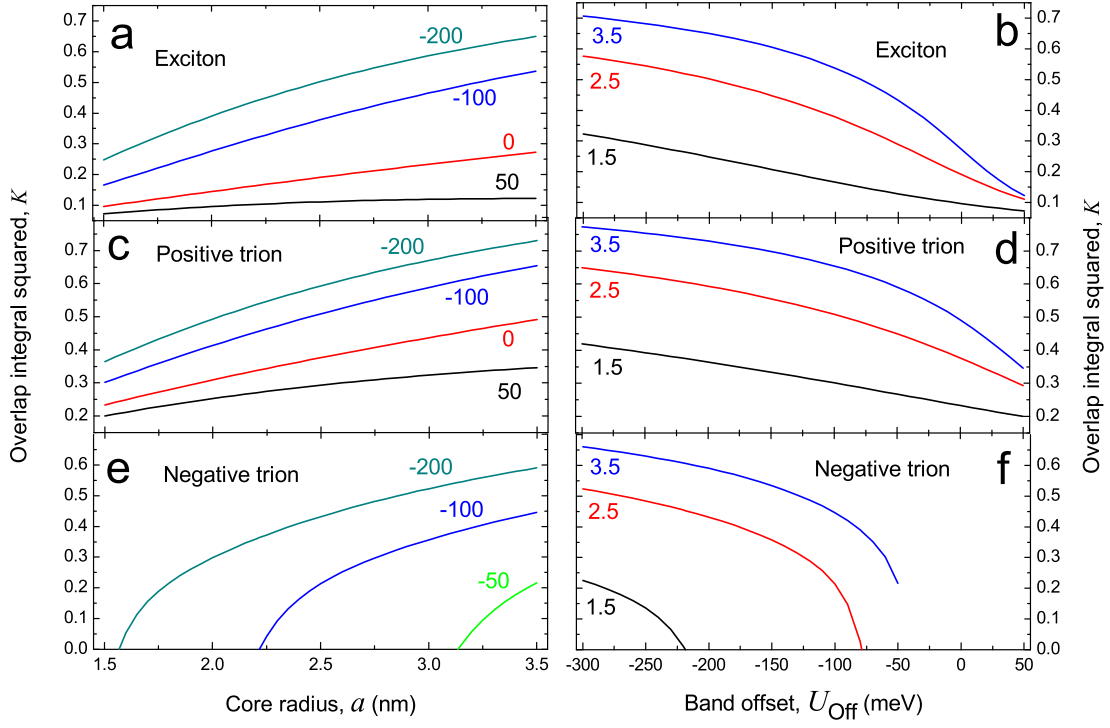, angle=0, width=1.0
\textwidth}
\end{center}
\vskip-0.7truecm
\caption{Dependence of the square of the overlap integral, $K$,  on the conduction band offset, $U_{\rm Off}$ (a), © and (e); and CdSe core radius, $a$ (b), (d), and (f),  calculated for an exciton (a) and (b); a positively charged trion © and  (d); and a negatively charged trion (e) and (f).  }
\label{Figovlp}
\end{figure}

The optical selection rule and the relative transition probabilities of the excitons in CdSe/CdS core shell NCs are described  by Eqs. (25)--(28) of Ref. \onlinecite{EfrosPRB96}, where the square of the overlap integral, $K$, between the electron and hole wave functions should be written in more general form
\begin{equation}
K_{ex} =  \left| \int dr r^2 R_e^{1}(r) R_0(r) \right|^2 \, ,
\end{equation}
 where $R_e^{1}$ is the electron radial function described in Eq. \eqref{Trial1}, with parameters $\alpha$ and $ \zeta$ that maximized  the exciton binding energy. In the  CdSe/CdS core shell NCs, where the electron is not localized in the CdSe core, $K_{ex}$ is a strong function of the band offset of the conduction band and of the core radius. In Figs. \ref{Figovlp}a and 
 \ref{Figovlp}b we show the dependence of the exciton  $K_{ex}$ on the core radius and the band offset, respectively.
\vskip-0.1truecm

The experimentally observed  increase of the the radiative decay time  by a factor of two in the temperature interval from 0 to 150 K \cite{BrovelliNatureComm11,RainoACSNano11,Benoit12} could be connected with a significant decrease  of the conduction band offset between CdSe and CdS shell layers. The decrease of the absolute value of the negative band offset decreases $K_{ex}$, as one can see in Fig. \ref{Figovlp} c, d and f.   The radiative decay time of the optically active excitons is inversely proportional to $K_{ex}$ and is directly proportional to the characteristic exciton decay time, as can be described by 
\begin{equation}
\frac{1}{\tau_{0}} = \frac{8 e^2 \omega n_r P^2 }{9m_0^2c^2\hbar}{\cal D}={{\cal D}\over \tau_0^i} ,
\end{equation}
where $n_r$ is the refractive index of the media, $c$ is the speed of light, $P=-i \langle S|\hat p_z|Z\rangle$  is the Kane transition matrix element,  $\omega$ is the light frequency, and ${\cal D}$ is the  depolarization factor, which takes into account the decrease of the magnitude of the light electric field when it penetrates into the NC. In CdSe structures $\tau_0^i$ should be on the order of 1.7\,ns.\cite{KlimovBook} This value was estimated using $n_r=1.5$, $\hbar\omega=2$\,meV and  $2P^2/m_0=19.0$\,eV.  The depolarization factor ${\cal D}$ generally depends on the NC shape and structure. In giant CdSe/CdS NCs the depolarization factor can be estimated using ${\cal D}=[3\epsilon/(2\epsilon+\epsilon_{\rm CdS})]^2$  for spherical CdS NCs, where $\epsilon_{\rm CdS}=5.35$ is the high frequency dielectric constant of CdS and $\epsilon=n_r^2=2.25$ is the dielectric constant of the media surrounding the NC. This results in  ${\cal D}\approx0.47$.

The radiative decay times of the Bright band-edge excitons has a complicated dependence on the exciton exchange energy $\eta_1$ and the hole level splitting $\Delta$.\cite{EfrosPRB96} The expression for the decay time is significantly simplified when  $\Delta\gg \eta_1$.  In this limit the radiative decay time, $\tau_{ex}$, of the first optically active exciton can be written as
\be
{1\over \tau_{ex}}={K_{ex}\over 2\tau_0}
\label{Exdecay}
\ee
This case is realized in giant CdSe/CdS NCs, due to a small exchange energy. From Fig.\ref{Figovlp} one can see  that  the overlap integral squared  decreases by a factor of 2--3 when CBO decreases from 50\, meV to 300 meV. This decrease of $K_{ex}$ is consistent with observed changes in the radiative decay time. One can see also from Fig.\ref{Figovlp}a that an increase of the core radius leads to a  strong localization of an electron in the core and increases $K_{ex}$ significantly. This in turn should shorten the radiative decay time.

\subsection{Positively charged trion transition probabilities}

The fine structure of the band-edge positive charge trions in NCs made of semiconductors with four-fold degenerate
 or quasi-degenerate valence  bands (like CdSe, CdS, InAs, CdTe and etc.)  generally consists of
four
  doubly degenerate  states  $E_{1/2}^\pm$, $E_{3/2}$ and $E_{1/2}$ and one
  four-fold degenerated state $E_{5/2}$, whose energies we described above. 
    All of these states are optically active and can decay radiatively with different decay
    times. 
This in principle could lead to a very complex dependence of the
decay time on the temperature as a result of the thermal population of
the different levels within the trion fine structure.

The resonant optical excitations of the positively charged trion in NCs with an extra hole and its radiative decay are determined by the probabilities of the  optical transitions  between  one of a hole state, $1S_{3/2}$, with $M=\pm1/2,\pm 3/2$ and the one of  the trion states, $Tr$ [$\pm 1/2^\pm$; $(3/2, \pm 3/2)$; $(5/2, \pm 3/2)$; $(5/2, \pm 5/2)$, and $(1/2, \pm 1/2)$]. These probabilities  are proportional to the following sum of the  matrix element squared:
\bea
T(M,Tr)&=&\sum_{\begin{array}{c} i,j=1,2; i \ne j  \end{array} }| \langle \Psi_{M}^h({\bm r}_{hi})
\delta ({\bm r}_{e} - {\bm r}_{hj}) | {\bm e} \hat {\bm p} | \Phi_{G,g}^{tr}( {\bm r}_{e},{\bm r}_{h1},  {\bm r}_{h2}) \rangle|^2 \nonumber\\
&=& 2|\langle \Psi_{M}^h({\bm r}_{h1})
\delta ({\bm r}_{e} - {\bm r}_{h2}) | {\bm e} \hat {\bm p} | \Phi_{G,g}^{tr}( {\bm r}_{e},{\bm r}_{h1},  {\bm r}_{h2}) \rangle|^2~.
\label{Ttr}
\eea
The sum takes into account the probability of an electron recombination with two holes ($j=1,2$). The sum can be replaced by a factor of two because the matrix elements of the transitions with different hole states are identical, due the symmetrization of the two-hole wave function.
  Here ${\bm e}$ is the polarization of the emitted or absorbed light,  and the momentum operator $\hat{\bm p}$ acts only on the valence band Bloch functions.  The calculation of $T(M,Tr)$ with the wave functions
is straightforward but cumbersome,  because the hexagonal axes of a NC ensemble are randomly oriented relative to the light  propagation of emission directions.
First, let us consider transition probabilities $T(M,Tr)$'s responsible for the radiative decay of the ground and first excited trion  state.

We will be considering here only the effect of linearly polarized light, which is sufficient for the description of the radiative decay. For linearly polarized light, ${\bm e} \hat {\bm p}$ in Eq. \eqref{Ttr} can be written as
  \be
{\bm e\hat{\bm p}}=e_z\hat{p}_z+\frac{1}{2}[e_-\hat{p}_++e_+\hat{p}_-]
\label{pol}
\ee
where $z$ is the direction of the hexagonal axis of the NC,
$e_{\pm}=e_x\pm ie_y$,
$\hat{p}_{\pm}=\hat{p}_x\pm i\hat{p}_y$, and $e_{x,y}$ and
$\hat{p}_{x,y}$ are the components of the polarization vector
and the momentum operator, respectively,  that are perpendicular to
the NC hexagonal axis.

In studied  CdSe/CdS core/shell NCs, the crystal field and shape asymmetry parameter $\Delta$ is much larger than the trion exchange parameter $\eta$. As a result the trion ground state is $1/2^-$.  Let us calculate the transition   probabilities $T(3/2,1/2^-;\pi)$ and  $T(-3/2,-1/2^-;\pi)$ between the trion ground state and the hole ground state $|M|=3/2$  stimulated by the linear polarized light.  
Substituting Eq. \eqref{pol} into Eq. \eqref{Ttr} and using the trion and hole wave-functions, after some straightforward but cumbersome calculations gives the following:
\begin{eqnarray}
  T(-3/2,-1/2^-;\pi)=T(3/2,1/2^-;\pi) =(B_2^{-}+B_0^{-})^2 \frac{K_{tr^+} P^2}{4} \sin^2\theta \, ,
\label{eq:44}
\end{eqnarray}
 where $\theta$ is the angle between the vector polarization of light  $\bm e$ and the hexagonal axis of the NC, and where the trion overlap integral squared
  $K_{tr^+}=\left| \int dr r^2 R_e^{Z=2}(r) R_0(r) \right|^2$ is determined now by the electron wave function $R_e^{Z=2}$ of the trion, which is calculated for the Coulomb potential created by two holes.  This radial function is described by  Eq. \eqref{Trial1}, with parameters $\alpha$ and $ \zeta$ that maximize  the trion binding energy.  The dependence of the overlap integral squared calculated for the positively charged trion is shown as functions of the CdSe core radius and CBO in Figs. 5 c and d.

Equation   \eqref{eq:44} describes the probability of the trion ground state radiative decay into the ground hole state.  There is nothing, however, that prevents the trion decay into the light hole excited states with $M=\pm 1/2$. The photon emitted during such decay will be smaller (by an energy of about $\Delta$) than the photon emitted during the ground-to-ground transition. This energy  will be absorbed by the hole. The probability of observing such a red "shake up" line is determined by the four independent transition probabilities:
\bea
   T(-1/2,-1/2^-;\pi)&=&T(1/2,1/2^-;\pi) =(B_2^{-}-B_0^{-})^2 \frac{K_{tr^+}P^2}{3} \cos^2\theta \, ,\nonumber\\
 T(1/2,-1/2^-;\pi)&=& T(-1/2,1/2^-;\pi) =(B_2^{-}-B_0^{-})^2 \frac{K_{tr^+}P^2}{12} \sin^2\theta \, ,
\eea
which are much smaller  than the probability of the observation of the main line,
because $(B_2^{-}-B_0^{-})^2 \ll (B_2^{-}+B_0^{-})^2$  for large $\Delta$.

Summing up the probabilities of the trion radiative decay  into the ground and excited hole states over all light polarizations, we obtain the radiative recombination rate of the trion ground state as
\begin{equation}
\frac{1}{\tau^{tr^+}_{1/2^-}} = \frac{K_{tr^+}}{2\tau_{0}}  \, ,
\label{eq:47}
\end{equation}
One can show that the uppermost trion excited state $1/2^{+}$ has the same radiative decay rate as the ground one ($\tau^{tr^+}_{1/2^+}=\tau^{tr^+}_{1/2^-}$) and they are different from the decay time of the first optically active exciton state only by the overlap integrals (see Eq. \eqref{Exdecay}).  Comparison of  the overlap integral squared  for excitons and positively charged trions in Fig. \ref{Figovlp} shows that in NCs with small conduction band offsets they became  much larger for the trions due to stronger attraction of electrons to the CdSe core.  Henceforth in this paper we  will use notations for the decay times, which are related to the notation of  corresponding levels of the positively charged trion:
$\tau^{tr^+}_{1/2^\pm}$, $\tau^{tr^+}_{3/2}$ and $\tau^{tr^+}_{1/2}$ and $\tau^{tr^+}_{5/2}$.

We find the probability of the transitions from the  excited trion states $(5/2,\pm 5/2)$ and $(5/2,\pm 3/2)$ to the hole $M=\pm3/2$ state to be
\be
  T(\pm 3/2,(5/2,\pm 5/2);\pi) = \frac{K_{tr^+}P^2}{6} \sin^2\theta\,,~
 T(\pm 3/2,(5/2,\pm 3/2);\pi) = \frac{2K_{tr^+}P^2}{3} \cos^2\theta \, .
\label{eq:50}
\ee
The transition to the hole ground state has an energy of about $\Delta$ larger then the main transition from the ground state.

The total transition probability from the trion $\pm 5/2$ states to the hole ground $\pm 1/2$ state is described by the sum of the two probabilities:
\be
T(1/2,(5/2,3/2);\pi)=T(-1/2,(5/2,-3/2);\pi))
  = \frac{K_{tr^+}P^2}{10} \sin^2\theta \, .
\label{eq:51}
\ee
In the same limit of large $\Delta$, transitions from the trion excited state to the hole excited state have nearly the same energy as the main transition from the ground to ground states.

When we average all possible crystal orientation and sum up Eqs. \eqref{eq:50} and \eqref{eq:51}, we can find the radiative recombination rate of the trion excited  state $E_{5/2}$:
\begin{equation}
\frac{1}{\tau_{5/2}^{tr^+}}  = \frac{3K_{tr^+}}{5\tau_0}  \, ,
\end{equation}
Similar calculations give us the
radiative rates for $E_{1/2}$ and $E_{3/2}$ excited trion states:
\begin{equation}
\frac{1}{\tau_{1/2}^{tr^+}} = {2K_{tr^+}\over 3\tau_0} \, , \quad \frac{1}{\tau_{3/2}^{tr^+}}={11K_{tr^+}\over 15\tau_0} \, .
\end{equation}

The situation is significantly  simplified in studied here CdSe/CdS core/shell NCs. Due to small  value of $\eta\ll \Delta$, the
    second trion level is eightfold degenerate  with an average decay rate of
\be {1\over \tau_2^{tr^+}}={1\over
4}\left(\frac{2}{\tau_{5/2}^{tr^+}}+\frac{1}{\tau_{1/2}^{tr^+}}+\frac{1}{\tau_{3/2}^{tr^+}}\right)=
{K_{tr^+}\over 2\tau_0}~, \ee which coincides with the decay rate of
the uppermost and lowest trion states $1/\tau^{tr^+}_{1/2^\pm}$ (see
Eq. \eqref{eq:47}). As a result, the positive trion radiative decay
time in CdSe/CdS core/shell NCs does not depend on
thermo-population of the fine level structure and  is temperature
independent.

\subsection{Negatively charged trion transition probabilities}

The radiative decay time of the negatively charged exciton leaves a resident electron in NCs. In CdSe/CdS core/shell  NCs, the wave function of the resident electron, $R_e^0$, is very different from the electron wave functions of the negatively charged trion, which is mainly controlled by the Coulomb interactions. The different size of the electron and hole wave functions  significantly decreases the overlap integral, which controls the rate of trion radiative decay.

The radiative decay and the resonant optical excitation of the negatively
charged trion are determined by the probabilities of the  optical transitions
between the resident "free" electron state, with spin projection
$S_z =\pm 1/2$ and radial wave function $R_e^0$, and one of the
trion states characterized by the hole momentum projection $M= \pm
3/2, \pm 1/2$.  These probabilities are proportional to
 \be 
T(S_z,M)= 2|\langle \Psi^e_{S_z}({\bm r}_1)  \delta ({\bm r}_{1} - {\bm
r}_{e2}) | {\bm e} \hat {\bm p} | \Psi_{M}^h({\bm r}_{h})
\Psi_{2e}({\bm r}_{e1},  {\bm r}_{e2}) \rangle|^2~,
\label{tr} 
\ee
where the factor 2 is due to the sum over electron coordinates.
The ground negative trion state is formed by the hole with $M= \pm
3/2$. The probability from the ground state is
\begin{eqnarray}
  T(-1/2,-3/2;\pi)=T(1/2,3/2;\pi) = \frac{K_{tr^{-}} P^2}{2} \sin^2\theta \,
  .
\label{eq:tr32}
\end{eqnarray}
Here, \be K_{tr^{-}}=|\int\int r_1^2dr_1 r_2^2 dr_2
R_e^0(r_1)R_{2e}(r_1,r_2)R_0(r_2) |^2 \, , 
\label{eq:61}
\ee 
where $R_{2e}$  is
the symmetrized two-electron wave function in Eq.(\ref{WF2e}),
$R_{0}$  is the wave functions of a single hole in
Eq.(\ref{Rfunction}), and ${R}_{e}^0$ is the wave function of a
single resident electron in Eq.(\ref{FreeWF}). 

Figures 5e and 5f show the dependence of the overlap integral squared, $K_{tr^-}$, for the negatively charged trion on the CdSe core radius, $a$,  and  the CBO, $|U_{\rm Off}|$. One can see from these figures that $K_{tr^-}$ always increases with $a$ and $|U_{\rm Off}|$.  At the same time
the figures show that for each $a$ there is critical $|U_{\rm Off}|$ and vise versa where $K_{tr^-}$ vanishes.  These critical points are connected with the delocalization of the resident electron into the CdS shell.  As we show above, the delocalization condition is described as
$w_0^{cr}= a\sqrt{2m_{\rm CdSe}|U_{\rm Off}}/\hbar\approx1.296$. At this condition the  spread of the electron wave function into the CdS shell vanishes the overlap integral. 
 
Taking a sum over all light polarizations in Eq.\eqref{eq:tr32},  we obtain the
radiative decay rate from the negative trion ground state as
 \be \frac{1}{\tau^{tr^-}_{3/2}} =
\frac{K_{tr^-}}{2\tau_0} \, . \ee 
As in the case of the positively charged trion, the radiative decay time of the negatively charged trion is different from the radiative decay time
of the first optically active exciton only by the overlap integral.
For the negative trion excited states with $M=\pm 1/2$, we
have following transition probabilities:
\begin{eqnarray}
  T(-1/2,-1/2;\pi)=T(1/2,1/2;\pi) = \frac{2K_{tr^-}P^2}{3} \cos^2\theta \,
  , \\
   T(1/2,-1/2;\pi)=T(-1/2,1/2;\pi) = \frac{K_{tr^-}P^2}{6} \sin^2\theta \,
  ,
\label{eq:tr12}
\end{eqnarray}
Taking a sum of the transition probabilities over all states and light polarizations, we obtain the
radiative decay rate of  the negative trion with $M=\pm 1/2$:
 \be \frac{1}{\tau^{tr^ -}_{1/2}} =
\frac{K_{tr^-}}{2\tau_0} \, . \ee 
The radiative decay time of the trion excited  state, $\tau^{tr^ -}_{1/2}$,  is the
same as one for the ground state: $\tau^{tr^ -}_{1/2}=\tau^{tr^ -}_{3/2}$.

\section{Discussion}

This paper has developed a theoretical  background  for qualitative and quantitative analyses of multiple experimental measurements conducted in
CdS/CdS core/thick shell NCs and CdSe/CdS dots-in-rods  nanostructures (NS). The major unusual properties of these NS are connected with the relatively small, and  temperature and strain dependent  conduction band offset,  which provides only weak confinement of electrons  in the CdSe core, while the large valence band offset in these NS leads to the strong confinement of holes.  As a result, the Coulomb potential created by strongly confined holes plays  an important role in electron confinement.

The optical properties of such NS should be considered to be in the {\it intermediate} confinement regime.\cite{Efros82}  
The calculation procedure  consists of several steps: (i) finding the energy spectra and wave function of strongly confined holes; (ii) calculation of the adiabatic Coulomb potential creating by the hole charge distribution; and, finally, (iii) calculation the energy spectra  of  the electron moving into the total potential created by the conduction band offset and by the Coulomb potential. In this paper, we realized this program assuming that the thickness of CdS shell is much larger than the  radius of electron localization around the CdSe core. The limit is satisfied  in "giant" CdSe/CdS core/shell NCs grown by the Los Alamos \cite{ChenJACS08,BrovelliNature11,HtoonNL10,ParkPRL11,Garcia-SantamariaNL09,BrovelliNatureComm11,GallandNatComm2012} and Deburter\cite{MahlerNM08,SpinicelliPRL09,Benoit12} groups.  The theory could be extended, however, to quantitative consideration of     
CdSe/CdS dots-in-rods  NS and/or to take into account the finite thickness of the CdS shell. 

Some effects of the finite CdS shell can be taken  into account  quantitatively  on the basis of the current calculations if the radius of electron localization around the CdSe core, $\zeta a$, is smaller than the CdS layer thickness.   For example, the negative trion ionization energy should be increased on the energy of the  first confined level in CdS NCs with the same shape and size as original CdSe/CdS core/shell NCs. The radiative decay  of the negative charged trion also does not vanish in NC with finite shell thickness when a delocalization condition is reached.  The time can be calculated by replacing $R_e^0$ in Eq. \eqref{eq:61} with the wave function of the the  first confined level in CdS NCs.

In the paper we  have also derived  general expressions for the radiative decay time of the band-edge trions in  various NCs. The calculations show that these times are proportional to $2\tau_0/K_{tr^\pm}$. The major difference in these times is connected with the square of the overlap integral between electron and hole wave functions in positively charged $K_{tr^+}$ and negatively charged $K_{tr^-}$ trions.  In CdSe/CdS core/thick shell NCs where the electron  exchange energy becomes smaller than the temperature, the radiative decay time of excitons is two times longer than $2\tau_0/K_{ex}$ given by Eq. \eqref{Exdecay} due to the equal population of the Dark and Bright exciton states.

A detailed comparison of our calculations  with experimental data on the optical properties of the CdSe/CdS core/shell NCs requires a knowledge of the temperature dependence of the conduction band offset.   This temperature dependence suggested first in Ref. \onlinecite{RainoACSNano11} is not universal, because it has a strain contribution \cite{Strain} and should depend on the sample preparation technique.  Nevertheless, the decrease of the radiative decay time by a factor of two observed in  Refs. \onlinecite{GallandNatComm2012} and \onlinecite{Benoit12} is consistent with  an increase of the overlap integral squared  with the CBO increased from 100 to 300 meV. The small splitting of the Dark/Bright exciton states
measured in  the giant CdSe/CdS NCs \cite{BrovelliNatureComm11} is also in qualitative agreement with our calculations.

In summary, we have developed a theory of the optical properties of CdSe/CdS core/thick shell NCs. The calculations takes into account the complex structure of the CdSe valence band and inter-particle Coulomb  and exchange interaction.  The theory describes the  fine structures and radiative decay time of excitons and  positively and negatively charged trions.  The results of our calculations are in qualitative agreement with experimental data available for these NCs. We believe that calculated energy spectra  of negatively and positively-charged trions   will explain suppression of nonradiative Auger processes observed in these structures. 

\section*{ACKNOWLEDGMENTS}
The authors thank J. Feldman for the critical reading of the manuscript.  A.S. acknowledges support of the Center for Advanced Solar 
Photophysics (CASP) an Energy Frontier Research Center founded by OBES, OS, U.S. DOE.  A.V.R. is grateful for the financial support
received from the Swiss National Science Foundation.
A. L. E. acknowledges support of
the Office of Naval Research and Alexander-von-Humboldt
Foundation.

\appendix*

\section{Normalization constant}

From the normalization condition for an electron bound to the singly or doubly charged core  
we obtain the following analytical expression for the normalization constant squared:
\begin{eqnarray*}
B^2 = {256 \left[2 \left(\alpha ^2 -1\right) \zeta +\alpha ^2 \mu \right]^2 \over
\alpha^3 \left[q\left( \mu, \alpha, \zeta\right) + 
\sqrt{2 \pi } \exp\left(2/\alpha^2\right)  \mathrm{erf}\left(\sqrt{2}/\alpha \right) p\left( \mu, \alpha, \zeta\right)\right]
}
\end{eqnarray*}

\noindent where $p\left( \mu, \alpha, \zeta\right)$ and $q\left( \mu, \alpha, \zeta\right)$ are given by
\begin{eqnarray*}
q\left( \mu, \alpha, \zeta\right) &=& 16 \alpha  \zeta^2
\left(4 - 55\alpha^2\right) + 256 \alpha  \zeta ^3 \left( 2 + 2 \zeta + \zeta^2\right) + 
16  \alpha^3 \zeta  \mu \left(27 \alpha^2 - 4\right)  \nonumber\\
&+& 4 \alpha ^5 \mu ^2\left(4-15 \alpha ^2\right)\nonumber \\
p\left( \mu, \alpha, \zeta\right) &=&
4 \zeta^2 \left( 16 - 56 \alpha^2 + 55 \alpha^4\right) + 
4 \alpha^2 \zeta \mu \left(40 \alpha^2 - 27 \alpha^4 - 16\right) \nonumber \\
&+& \alpha^4 \mu^2 \left(16 - 24\alpha^2 + 15 \alpha^4\right)
\end{eqnarray*}

\noindent and $\mathrm{erf}\left( x\right)$ is the error function defined as
\begin{eqnarray*}
\mathrm{erf}\left(x \right) = \frac{2}{\sqrt{\pi }}\int _0^x e^{-t^2}dt
\end{eqnarray*}


\end{document}